\documentclass[10pt,oneside]{article}

\usepackage{cmds11}
\usepackage{natbib}
\usepackage{graphicx}
\usepackage{times}
\usepackage[T1]{fontenc}
\usepackage{amsmath}
\usepackage{amssymb}
\usepackage{textcomp}
\usepackage{units}

\graphicspath{{./figs/}}

\title{ON THE RANGE OF 3D DISLOCATION PAIR CORRELATIONS}

\author{ {\sl F. F. Csikor$^{1,2}$, I. Groma$^{1}$,
    T. Hochrainer$^{3,4}$, D. Weygand$^{4}$, M. Zaiser$^{2}$} \\[5pt]
$^{1}$ Department of Materials Physics, E\"otv\"os University,
PO Box 32, H-1518 Budapest, Hungary \\
$^{2}$ Centre for Materials Science and Engineering, University of Edinburgh,
King's Buildings, Sanderson Building, Edinburgh EH9 3JL, UK \\
$^{3}$ Fraunhofer-Institut f\"ur Werkstoffmechanik IWM,
W\"ohlerstr. 11., 79108 Freiburg, Germany \\
$^{4}$ Institut f\"ur Zuverl\"assigkeit von Bauteilen und Systemen (izbs),
University of Karlsruhe, Kaiserstr. 12., 76131 Karlsruhe, Germany \\[5pt]
csikor@metal.elte.hu
}

\begin{document}

\maketitle

\paragraph {ABSTRACT:} Numerical studies of dislocation pair
correlations have played a central role in deriving a
continuum theory from the equations of motion of 2D dislocation
systems in a mathematically rigorous way. As part of an
effort to extend this theory into the full 3D dislocation problem,
3D dislocation pair correlations were studied with
discrete dislocation dynamics simulation. As a first approximation,
dislocations were modeled as uncharged curves in space (their
Burgers vectors were disregarded).
An inverse square decay with distance was found to describe the
numerically obtained pair correlations of the studied curve system.


\section{INTRODUCTION}
\label{sec:introduction}

Statistical mechanics studies of discrete 2D systems of straight,
parallel edge dislocations have led Groma and co-workers to a
rigorously derived prototype continuum model for bridging the micro-
and mesoscales of crystal plasticity.
Numerical calculation of dislocation pair correlation functions
played a crucial part in the construction of the theory
\cite{zaiser01,groma03}: i) The observed short
range nature of the correlations (which is nontrivial due to long
range elastic dislocation interactions) is a prerequisite for a local
theory to exist. ii) Pair correlation functions of homogeneous systems
are explicitly present in the theory in the flow stress term as well
as in a non-local diffusion-like term which enables the theory to
describe size effects.

In current efforts to generalize this theory to the full 3D
plasticity problem
\cite{elazab00,hochrainer07}, dislocation pair correlations are
expected to play a similar role as in the prototype 2D theory
\cite{zaiser06}. As there exist no theoretical predictions on the
functional form of 3D dislocation correlations, numerical simulation
is the obvious means to provide the necessary input for the continuum
theory. A pioneering 3D discrete dislocation dynamics (DDD) simulation
study
\cite{elazab07} considered the deformation of a symmetrically oriented
large bcc crystal with periodic boundary conditions and analyzed the
simulated dislocation network as a system of randomly distributed
curves (in mathematical terms, a stochastic fiber process),
disregarding the Burgers vector information. Unfortunately,
the single shot approach of this study did not provide enough
statistics to get smooth graphs as results and the authors did not
connect their numerical results to the crucial theoretical
questions mentioned above.

In the present paper, the results of a large number of statistically
equivalent DDD simulations are presented. As in \cite{elazab07},
the obtained dislocation configurations are analyzed as a system of
randomly distributed uncharged curves in space and the radial decay of the
pair correlation function of this curve system is studied.

\section{SIMULATIONS}
\label{sec:simulations}

The 3D dislocation configurations analyzed below were generated with
the DDD code described in \cite{weygand02,weygand05}. This code is
optimized for the small scale plasticity of finite, cuboid-shape fcc
single crystals and does not handle periodic boundary
conditions. In the model, dislocations are discretized into connected
straight segments which are allowed to glide. Junction formation
and dislocation annihilation upon contact are also included. Cross
slip was turned off in the present simulations to mimic low
temperature conditions.

In the following we present the simulated uniaxial tensile deformation
of (\unit[0.8]{\textmu{}m})$^{3}$ Al cubes up to an applied strain
$\epsilon_{\text{applied}}=0.67\%$ with an applied strain rate
\unit[5000]{$\text{s}^{-1}$}. The strain rate was chosen just below
the quasistatic limit, i.e.~above which the measured stress--strain
curves started to show enhanced hardening due to inertia effects. To
improve angular statistics, the crystals were oriented for [010]
symmetric multiple slip. The side faces of the specimens were
traction free. At the top and bottom faces displacement boundary
conditions were prescribed in the tensile direction whereas the in
plane components were traction free, too. The simulations were
started with 16 randomly positioned and oriented Frank--Read sources
of length \unit[0.22]{\textmu{}m} in each slip system,
resulting in an initial dislocation density
\unit[$\rho_{\text{initial}} = 8.3 \times 10^{13}$]{$\text{m}^{-2}$}.
%A typical snapshot near the
%end of a simulation run is displayed in fig.~\ref{fig:config}.

%\begin{figure}[tbh]
%  \centering
%  \includegraphics[height=4cm,clip]{config.eps}
%  \caption{Typical dislocation configuration at 0.67\% applied
%    strain. Red, green, blue, and cyan curves denote mobile
%    dislocations on the four \{111\} slip planes. Yellow lines
%    represent immobile Lomer locks created through dislocation
%    reactions. \label{fig:config}}
%\end{figure}

As mentioned in the Introduction,
for the present analysis the simulated dislocation configurations
were simplified by considering all
dislocations and junctions as simple curves in space (their
Burgers vectors were disregarded). We then calculated the radial
pair correlation function $g(r)$ of these curve systems (see
eq.~(\ref{eq:pc_calc}) for its definition) which was complicated by
three practical problems. i) To get smooth curves as results, we
needed to improve the
statistics by averaging over an ensemble of
identical simulations, different only in their random initial
configurations. In the following, an ensemble of 55 simulations is
analyzed which proved large enough to smooth out both the average
hardening behavior (see fig.~\ref{fig:hardening}) and the averaged
pair correlation functions (see below). ii) As can be seen in
fig.~\ref{fig:densityprofile}, the obtained
dislocation configurations were inhomogeneous with dislocation
depleted
zones near the sample surfaces. The thickness of these zones, around
\unit[0.2]{\textmu{}m}, was found to be roughly equal to the mean
dislocation--dislocation distance $1/\sqrt{\rho}$ (see also
fig.~\ref{fig:hardening} for volume averaged values of
$\rho$). Despite the different boundary conditions at the top and side
cube faces, no
visible differences were detected between these two types of
dislocation depleted zones. Therefore, in the following we restrict
our analysis to the central (\unit[0.4]{\textmu{}m})$^{3}$ sub-volume
which proved homogeneous to a good approximation
(see fig.~\ref{fig:densityprofile}). iii) The linear size of the
studied sub-volume was only four times
larger than the mean dislocation--dislocation distance
$1/\sqrt{\rho}$. This distorted each obtained pair correlation
function $g(r)$ already at distances $r$ where it was
still far from its asymptotic value. We corrected for this
effect by dividing the measured pair correlation functions with the
pair correlation function of a random point distribution in the
(\unit[0.4]{\textmu{}m})$^{3}$ sub-volume.

\begin{figure}[tbh]
  \centering
  \input{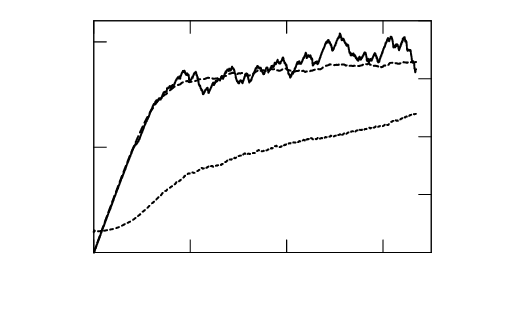}
  \caption{Solid line: typical simulated stress--strain curve. Dashed
    line: ensemble averaged stress--strain curve. Dotted line:
    evolution of the ensemble and volume averaged dislocation
    density. \label{fig:hardening}}
\end{figure}

\begin{figure}[tbh]
  \centering
  \input{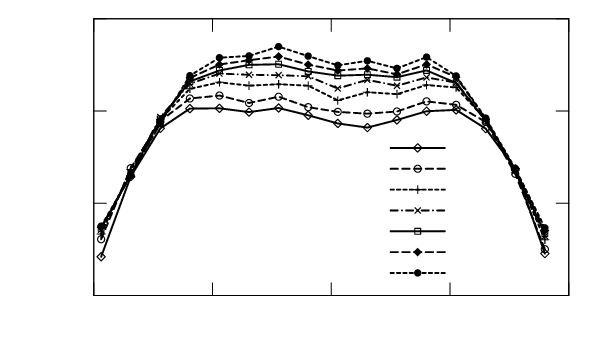}
  \caption{Evolution of the ensemble averaged density profile,
    averaged over the tensile and one of the perpendicular
    directions. \label{fig:densityprofile}}
\end{figure}

\section{DISCUSSION}
\label{sec:discussion}

\begin{figure}[tbh]
  \centering
  \input{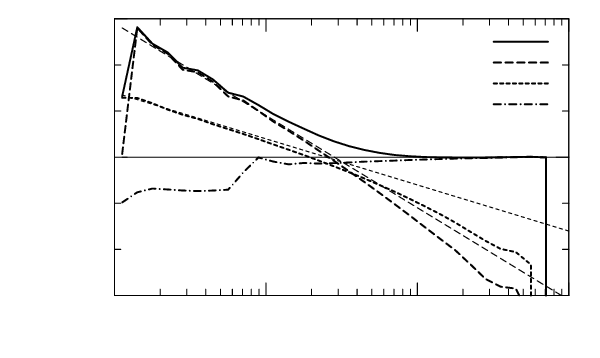}
  \caption{Decomposition of the pair correlation function $g(r)$ at
    0.67\% applied strain (see
    text for details). The $r \to 0$ asymptotes of $g_{\text{self}}(r)
    \propto r^{-2}$ and $g_{\text{ints}}(r) \propto r^{-1}$, and the
    $r \to \infty$ asymptotes of $g(r) \to 1$ and $g_{\text{pair}}(r)
    \to 1$ are also indicated. \label{fig:pc_all}}
\end{figure}

The radial pair correlation function of a random system of curves can
be defined as
\begin{equation}
  \label{eq:pc_calc}
  g(r) = \left( \frac{d}{dr} L(r) \right) / 4 \pi r^2 \rho
\end{equation}
\cite{stoyan86} where $L(r)$ is the mean total line length in a sphere
of radius $r$ centered at a
``typical'' curve point and $\rho$ means the line density
(total length in a unit volume). In this context, ``typical'' means
that a random choice is made in a way that every point on the curves
has the same chance to be chosen. The numerical method to compute
$g(r)$ directly followed its definition: i) The entire
dislocation network (both mobile dislocations and junctions) was
divided into line segments of length \unit[0.5]{nm}. ii) For every
line segment the total curve length was calculated in a sphere of
radius $r$. iii) The curve length values were averaged over all
line segments and then substituted into eq.~(\ref{eq:pc_calc}) as the
numerical estimate for $L(r)$. The results of such a calculation at
$\epsilon_{\text{applied}}=0.67\%$ are displayed in
fig.~\ref{fig:pc_all} (solid line). (It was found that
$g(r)$ depends only slightly on $\epsilon_{\text{applied}}$ and
it seems to saturate with increasing
$\epsilon_{\text{applied}}$. Further simulations to confirm this up to
$\epsilon_{\text{applied}} = 1.3\%$ are in progress). As can be seen
in fig.~\ref{fig:pc_all}, $g(r)$ diverges as
$r^{-2}$ as $r \to 0$ and converges to $1$ as $r \to \infty$. The main
goal of this study is to establish the rate of this
convergence. Before this can be done, however, $g(r)$
needs further analysis for two reasons. i)
The contiguity of dislocation lines causes strong correlations at
small $r$ values. These need to be separated from $g(r)$
as only the correlations between different dislocation lines are
interesting for theory \cite{zaiser06}. ii) In the simulations,
dislocation lines are discretized into linear segments of length
\unit[13--66]{nm} which might also distort the numerically obtained
$g(r)$ data. In the following, we analyze $g(r)$ along these problems.

The connected nature of dislocation lines causes two types of strong
correlations at small $r$: i) self correlations of
smooth dislocation curves $g_{\text{self}}(r)$ and ii)
correlations between different dislocations near their
intersections (e.g.~arms going out of junctions)
$g_{\text{ints}}(r)$.
First we analytically calculate the $r \to 0$ asymptotes of these two
contributions utilizing eq.~(\ref{eq:pc_calc}), then separate them
from the numerical $g(r)$ data at small $r$ values to estimate their
impact.

As $r \to 0$, $g_{\text{self}}(r)$ is equivalent to the pair
correlation function of randomly distributed straight lines with
a linear density
$\rho$. We only need that part of the correlation function where a
line correlates with itself. From eq.~(\ref{eq:pc_calc}) it readily
follows that
\begin{equation}
  \label{eq:pc_self}
  g_{\text{self}}(r) \to 1 / 2 \pi \rho r^{2} \text{ as } r \to 0
\end{equation}
(the same result was also derived in \cite{stoyan86}).

The $r \to 0$ asymptote of $g_{\text{ints}}(r)$ can be modeled with
``corners'' homogeneously distributed in space. In a ``corner'' two
arms of length $l$ meet at an angle $\varphi$ and the arms have a
linear density $\rho$. For the $r \to 0$ asymptote of
$g_{\text{ints}}$, only that part of the
correlation function is interesting where an arm correlates with the
arm attached to it. Straightforward calculation from
eq.~(\ref{eq:pc_calc}) yields
\begin{equation}
  \label{eq:pc_ints}
  g_{\text{ints}}(r) \to \frac{1}{4 \pi} \frac{1}{l \rho}
  \frac{\pi - \varphi}{\sin(\varphi)} \frac{1}{r} \text{ as } r \to 0.
\end{equation}
Note that in the case of simulated dislocation configurations,
averaged $l$ and $\varphi$ values appear in eq.~(\ref{eq:pc_ints}) but
this does not affect the $r^{-1}$ character of $g_{\text{ints}}(r)$ at
small $r$.

The next step is assessing the relative contributions of
$g_{\text{self}}(r)$ and $g_{\text{ints}}(r)$ to the numerically
calculated pair correlation function $g(r)$. To this end, we separated
$g(r)$ into three terms utilizing the fact that
the simulation code stores dislocation segments organized into
loops. The first term, $g_{\text{self}}(r)$, was
computed as the correlation function of segment pairs residing on the
same loop. The second one, $g_{\text{ints}}(r)$, was calculated
as the pair correlation of segment pairs on different loops which
touch each other. The last term, $g_{\text{pair}}(r)$,
was defined simply as
$g(r)-g_{\text{self}}(r)-g_{\text{ints}}(r)$. Fig.~\ref{fig:pc_all}
displays these three contributions and the total pair correlation
function $g(r)$ at $\epsilon_{\text{applied}}=0.67\%$. Numerical fits
of the
$r^{-2}$ asymptote of $g_{\text{self}}(r)$ and the $r^{-1}$ asymptote
of $g_{\text{ints}}(r)$ at $r \to 0$ are also displayed. Both fit well
for $r \lesssim 10\text{ nm}$, consistently with the minimum
discretization segment length \unit[13]{nm}, and diverge from the
numerical curves at larger
distances. Note that $g_{\text{pair}}(r)$ shows no singularity at
$r\to{}0$, i.e.~all connectivity related pair correlation
terms were successfully separated. Note also
that $g_{\text{pair}}(r) < 1$ (is anticorrelated) and that, to a
smaller extent, even $g(r) < 1$ for $r > 0.1 \text{ \textmu{}m}$
(not visible on the graph). The latter is clearly a finite size effect
caused by the independence of $\int_{0}^{\infty} g(r) r^{2} dr$
from $g(r)$, that $g(r)=1$ is a valid pair correlation function,
and that $g(r) \gg 1$ as $r \to 0$. The authors took the liberty
to manually compensate for this effect by adding a small constant $c
\ll 1$ to $g(r)$ when analyzing its decay to $1$.

\begin{figure}[tbh]
  \centering
  \input{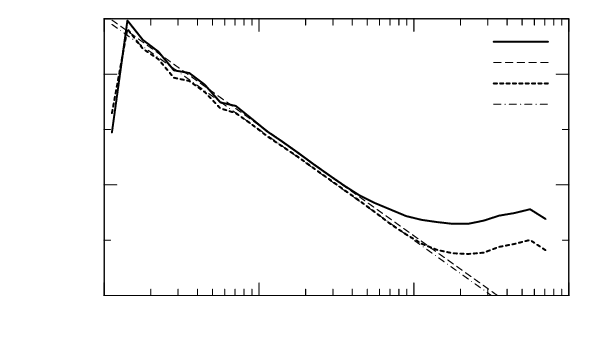}
  \caption{Evolution of the $r \to \infty$ decay of the pair
    correlation function $g(r)$. \label{fig:pc_totl_evolution}}
\end{figure}

We finally study the decay of $g(r)$ towards the uncorrelated value
$1$ as $r \to \infty$.
Fig.~\ref{fig:pc_totl_evolution} depicts $g(r)-1+c$ for
$\epsilon_{\text{applied}}=0$ and $\epsilon_{\text{applied}}=0.67\%$.
With $c=0$ (not shown in the figure), a $g(r)-1 \propto
r^{-2}$ decay can be seen up to $r \approx 20 \text{ nm}$ for
$\epsilon_{\text{applied}}=0$ and $r \approx 30 \text{ nm}$ for
$\epsilon_{\text{applied}}=0.67\%$ before $g(r)$ submerges
$1$. However, the authors think such an anticorrelated $g(r)$
unlikely. Instead, by carefully
adjusting $c$, an amazingly good fit to $r^{-2}$ can be achieved:
at $\epsilon_{\text{applied}}=0$ with $c=0.25$ up to $r=50\text{ nm}$
and at $\epsilon_{\text{applied}}=0.67\%$ with $c=0.08$ up to
$r=100\text{ nm}$ ($1.5$ times the discretization segment length
\unit[66]{nm}; see fig.~\ref{fig:pc_totl_evolution}). It is clear from
fig.~\ref{fig:pc_all} that at
$\epsilon_{\text{applied}}=0.67\%$ this $r^{-2}$ decay is not plainly
a result of the same-loop correlation $g_{\text{self}}(r)$
(cf.~eq.~(\ref{eq:pc_self})) as $g_{\text{self}}(r)$ is
$1.5$ orders of magnitude smaller than $g(r)$
at $r=100\text{ nm}$. It has also nothing to do with individual
straight segments as they are shorter than
\unit[66]{nm}. Therefore, the observed $r^{-2}$ decay of $g(r)$ is a
collective effect at larger distances. This is notable from the point
of view that the upper limit of this decay in our simulations,
$r=100\text{ nm}$, is still in the range of the mean
dislocation--dislocation distance $1/\sqrt{\rho}$ (see
fig.~\ref{fig:hardening}).

In summary the results indicate a gradual extension of an
$r^{-2}$ correlated zone from small $r$ values towards larger ones as
the deformation proceeds. A new simulation effort is ongoing to reach
$\epsilon_{\text{applied}}=1.3\%$ for further confirmation.

In a final note, a pair correlation function $\propto r^{-2}$
should be proportional to $\rho^{-1}$ for dimensional reasons. This is
consistent with the numerical results in
fig.~\ref{fig:pc_totl_evolution} (see also fig.~\ref{fig:hardening}
for $\rho$ values).

\section{CONCLUSIONS}
\label{sec:summary}

The correlation properties of 3D many-dislocation systems were
studied with discrete dislocation dynamics simulation in
symmetrically oriented fcc crystals deformed in uniaxial
tension. As a first
approximation, dislocation configurations were analyzed as
uncharged curve systems in space and the evolution of the
corresponding radial pair correlation function was studied. The
results indicate the gradual appearance of a pair
correlation function $\rho^{-1} r^{-2}$ from small $r$ towards larger
ones as the
deformation proceeds. Interestingly, this decay is identical to that
of 2D systems of infinite, parallel edge dislocations relaxed in
multiple slip \cite{unpub}. Moreover, substituting the numerical
result $\rho^{-1} / |\mathbf{r} - \mathbf{r}'|^{2}$ into eqs.~(10--11)
in \cite{zaiser06} as the double angular integral of $d^{\text{pair}}$
suggests that the dimensionless constants denoted by $\alpha$ and
$D(\theta)$ in \cite{zaiser06} have at most a logarithmic divergence
with system size. Whether this divergence is cancelled by the combined
angular dependence of $g$ and the dislocation pair interaction $\tau$,
enabling finite $\alpha$ and $D(\theta)$ values and
thus a local 3D dislocation continuum theory, is the
objective of subsequent numerical work.

% The following steps are planned to extend this work. We plan to study
% the angular
% properties of pair correlation functions (similar to
% \cite{elazab07}). Furthermore, taking Burgers vectors into account
% would allow studying the pair correlations of the Nye dislocation
% density tensor which
% is explicitly present in 3D continuum theories of dislocation
% systems. Finally, relaxation of the applied stress at certain applied
% strain values would allow studying the effect of dipole polarization
% on the correlation properties.

\section*{ACKNOWLEDGMENT}

Financial support of the European Community's Human Potential
Programme under Contract No.\ MRTN-CT-2003-504634 [SizeDepEn] is
gratefully acknowledged.

\bibliographystyle{unsrt}
%\bibliography{cmds11_csikor}

\end{document}